\documentclass[11pt]{article}
\usepackage{epsfig,latexsym}

\def\lapprox{\mathrel{\mathop  {\hbox{\lower0.5ex\hbox{$\sim$}
\kern-1.1em\lower-0.7ex\hbox{$<$}}}}}
\def\gapprox{\mathrel{\mathop  {\hbox{\lower0.5ex\hbox{$\sim$}
\kern-1.1em\lower-0.7ex\hbox{$>$}}}}}

\begin{document}

\begin{titlepage}
\title{The response of primordial abundances 
to a general modification of $G_{\rm N}$ and/or of 
the early universe expansion rate}
\author{C. Bambi$^{a,b,}$\footnote{E-mail:
bambi@fe.infn.it}, M. Giannotti$^{a,b,}$\footnote{ E-mail:
giannotti@fe.infn.it } and F.L. Villante$^{a,b,}$\footnote{
E-mail:
villante@fe.infn.it }}

\maketitle
\begin{center}
$^{a}$Istituto Nazionale di Fisica Nucleare, Sezione di Ferrara,
       I-44100 Ferrara, Italy\\
$^{b}$Dipartimento di Fisica dell'Universit\`a di Ferrara,
       I-44100 Ferrara, Italy
\end{center}

\vspace{0.5cm}

\begin{abstract}
We discuss the effects 
of a possible time variation of the Newton constant $G_{\rm N}$ on 
light elements production in Big Bang Nucleosyntesis (BBN).
We provide analytical estimates for the dependence of 
primordial abundances on the value of the Newton constant during BBN.
The accuracy of these estimates is then tested by numerical methods.
Moreover, we determine numerically the response of each element to an arbitrary time-dependent modification
of the early universe expansion rate. Finally, we determine the bounds
on possible variations of $G_{\rm N}$ which can be obtained from the 
comparison of theoretical predictions and observational data.

\end{abstract}

\end{titlepage}

\section{Introduction}
The idea that fundamental constants may vary with time dates back to
Dirac \cite{dirac}. Even if, at present, there is no robust experimental
evidence in favor of this possibility, this idea continues to be widely discussed,
since many extensions of the standard theories (e.g. superstring theories,
scalar-tensor theories of gravitation, etc.) predict 
a variation of "fundamental constants" with time.

On a pure theoretical level, the space-time dependence of the
fundamental parameters\footnote{
In this context, the term "fundamental parameters" is more appropriate than 
"fundamental constants".}
is forbidden by the 
Strong Equivalence Principle (SEP), and in particular by the
statement of Local Position Invariance (LPI) (see e.g. \cite{will}). On the other hand, 
the weakest form of the Equivalence Principle, 
the so-called Einstein Equivalence Principle (EEP)
-- which is the essence of the geometrical theory of gravitation --
refers (in its LPI) 
only to non-gravitational physics and, thus,
allows the Newton constant $G_{\rm N}$ to be time-dependent. This is what 
happens, e.g., in the Brans-Dicke, or more generally in the 
(multi-) scalar-tensor theories of gravity.  

The above argument suggests that $G_{\rm N}$
has a special role in the subject of time variation of the fundamental
parameters.
A dependence of $G_{\rm N}$ on time is a symptom of the violation 
of the SEP, but not necessarily of the EEP, whereas the non-constancy
of the other constants, like the electroweak or strong coupling constants, 
necessarily represents a violation of the equivalence principle in both its forms.

In this paper, we discuss the effects of time variations of $G_{\rm N}$
on the light element production in Big Bang Nucleosynthesis (BBN),
completing and extending the results of previous analysis on the subject 
(see e.g. \cite{rothman, accetta, krauss}).
BBN is evidently a good probe of a possible time variation of $G_{\rm N}$, 
since it is the earliest event in the history of the universe for which we can obtain
solid and well-testable predictions. Even a weak (or very peculiar) time 
dependence, which gives no observable effects in high accuracy experiments
performed at the present epoch, could give sizable effects when translated 
over cosmological time scales. BBN, however, is a complex phenomenon, since
each element responds in its own way to a modification of $G_{\rm N}$. 
We devote particular attention to this point,
introducing suitable response functions which relate the abundance of each element
to an arbitrary time-dependent modification of the early universe expansion rate.
 
The plan of the paper is the following: In the next section, we discuss
the role of $G_{\rm N}$ in BBN and we derive, analytically, the dependence 
of the primordial abundances on the value of the Newton constant at the key
epochs. In section 3, we calculate numerically
the response functions,
emphasizing that different elements are sensitive to the value of the Newton 
constant at slightly different times. 
In section 4, we discuss the bounds on $G_{\rm N}$ variations that can be
obtained from 
the comparison of theoretical predictions with the observational data for light elements primordial abundances. We summarize our results in section 5.

\section{The role of $G_{\rm N}$ in BBN}

The production of light elements (namely $^{2}{\rm H}$, $^{3}{\rm He}$, 
$^{4}{\rm He}$ and $^{7}{\rm Li}$) in
BBN is the result of the efficiency of
weak reactions ($p+e\leftrightarrow n+\nu_{\rm e}$ and related processes) and 
nuclear reactions (which build light nuclei from neutrons and protons) 
in the expanding universe.
The value of the gravitational constant determines the expansion rate
of the universe 
and thus, in turn, the 
relevant time scales for the above processes. As a consequence, if we assume
that the gravitational constant at time of BBN
is different 
from its present value, 
this translates into a variation 
of light element abundances with respect to standard BBN predictions. 

The above argument clarifies in simple terms the role of $G_{\rm N}$ 
in BBN. In order to be more quantitative, one needs, as a first step, to identify
the key epochs for light element production with respect to possible 
$G_{\rm N}$ variations. As we shall see, the relevant periods are those 
during which the weak reaction rates and/or the nuclear reaction rates 
are not vanishing nor exceeding
the universe expansion rate. Essentially they are the weak-interaction 
``freeze-out'' epoch (which occurs at the
temperature $T_{\rm f}\sim 0.8$ MeV) and the
``deuterium bottleneck'' epoch 
(which corresponds to $T_{\rm d}\sim 0.08$ MeV) \cite{kolb,bernstein,mukhanov,dimopoulos}.
In the following, we use $G_{\rm N,f}$ and $G_{\rm N,d}$ to indicate the
value of the Newton constant during these periods, while we use $G_{\rm N,0}$
to indicate the present value.

In order to estimate the dependence of the various elemental abundances
on $G_{\rm N,f}$ and $G_{\rm N,d}$,  it is necessary to 
quickly review the basic physical mechanisms responsible for light
element production.
When $T \gg T_{\rm f}$, the rate of weak processes which interchange
neutrons and protons, $\Gamma_{\rm W} \sim G_{\rm F}^2 T^5$ ,
is large with respect to the expansion rate of the universe:
\footnote{Here and in the following, we use a natural system of units in which 
$\hbar=c=k_{\rm B}=1$.}
\begin{equation}\label{H}
H = 1.66\, \sqrt{g_{*}G_{\rm N}} \,T^{2}~,
\end{equation}
where $g_{*}$ counts the total number of relativistic degrees of freedom\footnote{
Strictly speaking, equation (\ref{H}) is derived in the context of General 
Relativity in which $G_{\rm N}$ is constant.
In any extension of the standard theory, one has additional terms related to
time derivatives of $G_{\rm N}$. In this paper, we make the usual assumption
that $G_{\rm N}$ is slowly varying (with respect to the early universe 
expansion rate) which implies that these extra terms are negligible.}. 
As a consequence, neutrons and protons are in chemical equilibrium
and the neutron abundance $X_{\rm n} = n_{\rm n}/n_{\rm B}$, 
defined as the ratio of neutron to baryon  densities, 
is simply given by $X_{\rm n}(T)=[1+\exp(\Delta m/T)]^{-1}$,
where $\Delta m\simeq 1.29$ MeV is the neutron-proton 
mass difference.

For $T \le T_{\rm f}$ the weak reaction rate 
drops below the Hubble expansion rate, 
the neutron abundance freezes out 
at the equilibrium value $X_{\rm n}(T_{\rm f})$ 
and it then evolves only due to the neutron decay:
$X_{\rm n}(t)\simeq X_{\rm n}(T_{\rm f})\exp(-t/\tau)$, 
where $\tau=885.7$~s is the neutron lifetime. 
The ``freeze-out'' temperature is basically
determined by the condition $\Gamma_{\rm W}(T_{\rm f})/H(T_{\rm f})\simeq 1$ and,
clearly, depends on the value of the gravitational constant. 
One obtains:~\footnote{See \cite{weinberg, dimopoulos} for the precise numerical calculation 
of the total weak rate $\Gamma_{\rm W}$.}
\begin{equation}\label{freeze-out}
	T_{\rm f}=0.784 \left(\frac{G_{\rm N,f}}{G_{{\rm N},0}}\right)^{1/6}{\rm MeV}~,
\end{equation}
where $G_{\rm N,f}$ is the value of the Newton constant during the freeze-out, 
i.e. when the temperature of the universe is $0.2 \le T \le 2$ MeV (see next section). 
The larger is $G_{\rm N,f}$, the earlier is 
the freeze-out of the neutron abundance, at an higher value, and
the larger is the $^{4}{\rm He}$ abundance produced in BBN.

Light element production in BBN occurs through a sequence
of two body reactions, such as $p(n,\gamma)^{2}{\rm H}$, 
$^{2}{\rm H}(d,n)^{3}{\rm He}$,
$^{2}{\rm H}(d,p)^{3}{\rm H}$, 
$^{3}{\rm He}(p,\gamma)^{4}{\rm He}$, etc. 
Deuterium has to be produced in appreciable quantity before 
the other reactions can proceed at all. 
Nucleosynthesis effectively begins when the rate of deuterium processing
through $^{2}{\rm H}(d,n)^{3}{\rm He}$ and $^{2}{\rm H}(d,p)^{3}{\rm H}$
reactions
becomes comparable with the expansion rate of the universe.
By imposing this condition one finds the ``deuterium bottleneck'' temperature 
which, in standard BBN, is given by: 
$T_{\rm d}=0.08 (1+0.16 \log (\eta/10^{-10}))~{\rm MeV}$ \cite{mukhanov}.

After nucleosynthesis has started, light nuclei ($^{2}{\rm H}$, $^{3}{\rm He}$, 
$^{4}{\rm He}$ and $^{7}{\rm Li}$) are quickly produced. 
The $^{4}{\rm He}$ abundance is basically determined by the 
total number of neutrons that survive till the onset of nucleosynthesis,
since nearly all available neutrons are finally captured in 
$^{4}{\rm He}$ nuclei.
 The synthesized elemental abundances of $^{2}{\rm H}$, $^{3}{\rm He}$
 and $^{7}{\rm Li}$ are, instead, the result of the complex interplay 
 of the various nuclear reactions
efficient during and  after the d-bottleneck. 
The value of the Newton constant during this period, $G_{\rm N,d}$, 
clearly plays a relevant role both for $^{4}{\rm He}$ and other 
elements production. 

The situation with $^{4}{\rm He}$ is particularly simple.
The primordial $^{4}{\rm He}$ mass fraction 
is approximatively given by $Y_{4}\sim 2 X_{\rm n}(t_{\rm d})\simeq
2 X_{\rm n}(T_{\rm f})\exp(-t_{\rm d}/\tau)$, where
$t_{\rm d}$ is the age of the universe at the d-bottleneck.
Neglecting the weak dependence of the
temperature $T_{\rm d}$ on $G_{\rm N}$, one has:
\begin{equation}\label{d_bottleneck}
	t_{\rm d}\sim 206
	\left(\frac{G_{\rm N,d}}{G_{{\rm N},0}}\right)^{-1/2} (1-0.32 
	\log (\eta/10^{-10}))~{\rm sec},
\end{equation}
where $G_{\rm N,d}$ is the value of the Newton constant when the 
temperature of the universe is $0.02 \le T \le 0.2~{\rm Mev}$ (see next section),
i.e. when $^{4}{\rm He}$ and the other elements are effectively synthesized.   
By using this formula and considering eq.~(\ref{freeze-out}), one is able to estimate
the dependence of $Y_{4}$ from $G_{\rm N,f}$, $G_{\rm N,d}$ 
and $\eta$. One obtains:
\begin{equation} \label{Yp}
  \delta  Y_{4}\equiv\frac{\Delta Y_{4}}{Y_{4}}=
  0.23~\delta G_{\rm N,f}+0.09~\delta G_{\rm N,d}
  + 0.07~\log (\eta/\eta_{\rm CMB})
\end{equation}
where $\delta G_{\rm N,i}$ (with ${\rm i} = { \rm f,d}$) represents 
the fractional variation of the Newton constant at a given
epoch with respect to its present value:
\begin{equation}
\delta G_{\rm N,i} = \frac{ G_{\rm N,i}-G_{\rm N,0}}{G_{\rm N,0}} ~,
\end{equation}
and $\log(\eta/\eta_{\rm CMB})$
is the logarithmic variation of the baryon to photon ratio with respect to
the value $\eta_{\rm CMB} = 6.14\cdot10^{-10}$ presently favored 
by cosmic microwave background (CMB) \cite{spergel} and deuterium data \cite{d7}.

The situation with $^{2}{\rm H}$, $^{3}{\rm He}$ and $^{7}{\rm Li}$
is slightly more complicated. In principle, one has to integrate the
rate equations, which can be written formally as: 
\begin{equation}\label{rates}
 \frac{{\rm d}Y_i}{{\rm d}t} \propto \eta\,n_{\gamma}\, \sum_{+,-} Y \times Y \times 
  \langle\sigma v\rangle_T\ ,
\end{equation}
where $Y_{i}$ indicate the abundance of a given element, 
the sum runs over the relevant source $(+)$ and sink $(-)$
terms, and $\langle \sigma v\rangle_T$ are the thermally-averaged
reaction rates.\footnote{
In writing eqs.~(\ref{rates}) and (\ref{rates2}), we implicitly assumed that
only two-baryon reactions are relevant (neglecting reactions such as 
$p+e\rightarrow n+\nu$ and related processes or $d+\gamma\rightarrow n+p$). 
This is reasonable after deuterium bottleneck and provides a 
good framework to discuss the abundances of $^{2}{\rm H}$, 
$^{3}{\rm He}$ and $^{7}{\rm Li}$, whose abundance is essentially
established after the d-bottleneck. On the contrary, this is clearly
not adeguate to discuss the synthesis of $^{4}{\rm He}$.} 
In order to estimate the role of $G_{\rm N}$,
one can simply note that, since the temperature of the 
universe evolves as $dT/dt\propto-T^3\sqrt{G_{\rm N}}$, the above 
equation can be rewritten as:
\begin{equation}\label{rates2}
 \frac{{\rm d}Y_i}{{\rm d}T} \propto -\frac{\eta}{G_{\rm N,d}^{1/2}}
  \,\frac{n_{\gamma}}{T^3}\, \sum_{+,-}  Y \times Y \times \langle\sigma v\rangle_T\ ,
\end{equation}
which shows that $^{2}{\rm H}$, $^{3}{\rm He}$ and $^{7}{\rm Li}$
depend on $\eta$ and $G_{\rm N}$ essentially through the 
combination $\eta/G_{\rm N,d}^{1/2}$. 
This suggests that the synthesized elemental
abundances of ${\rm D}/{\rm H}$, $^3{\rm He}/{\rm H}$, and $^7{\rm
Li}/{\rm H}$ (indicated, in the following, with $Y_2$, $Y_3$, and $Y_7$ respectively)
for an arbitrary value of $G_{\rm N,d}$ 
can be related to the standard case ($G_{\rm N,d}=G_{\rm N,0}$) 
through an appropriate rescaling in $\eta$:
\begin{equation}\label{shift}
Y_{i}(\eta,G_{\rm N,d}) = Y_{i}\left(\eta (G_{\rm N,0}/G_{\rm N,d})^{1/2}  ,G_{\rm N,0}\right)~.
\end{equation}
The previous equation, linearized, implies that:
\begin{equation}\label{smallshift}
\delta Y_{i} \equiv \frac{\Delta Y_{i}}{Y_{i}} =  
\gamma_{i}(\eta_{0},G_{\rm N,0}) \left( \Delta \log (\eta) - 0.22 
\,\delta G_{\rm N,d} \right)\end{equation} 
where $\eta_{0}$ is an arbitrary pivot point, 
$\gamma_{i}=(1/Y_{i})\,\partial Y_{i} / \partial \log \eta$
and we have assumed that $\Delta \log(\eta)=\log(\eta/\eta_{0})$ and $\delta G_{\rm N,d}$
are small. We remark that, according to eqs.~(\ref{shift},\ref{smallshift}), fixed values of the abundances 
correspond in the plane $(\log\eta,\delta G_{\rm N,d})$
to straight parallel lines (with slope $\sim 1/0.22$).

The above argument is accurate enough to describe $^2{\rm H}$ and $^3{\rm He}$
abundances. However, it can be slightly improved in order to predict correctly the 
behavior of  $^7{\rm Li}$. In order to produce $^{7}{\rm Li}$, one has to 
use $^{4}{\rm He}$ as a target, whose abundance is strongly dependent on 
the value of the Newton constant. This clearly 
introduce an extra dependence on $G_{\rm N,f}$ and 
$G_{\rm N,d}$. One expects an extra factor in eq.~(\ref{shift})
proportional to $Y_{4}$ which, linearized, gives:
\begin{equation}\label{li7}
\delta Y_{7} =
 ( 0.23~\delta G_{\rm N,f}+0.09~\delta G_{\rm N,d})+
 \gamma_{7}(\eta_{0}) \left( \Delta \log (\eta) -0.22 \,\delta G_{\rm N,d} \right) 
\end{equation}
where the first terms in the r.h.s are obtained from eq.~(\ref{Yp}) and we neglected 
the weak dependence of $Y_{4}$ on the baryon to photon ratio $\eta$.

The analytical results discussed above allow us to understand 
the relevance of the various physical mechanisms in light elements production
and to obtain simple quantitative relations (eq.~(\ref{Yp}) for $^{4}{\rm He}$, 
eq.~(\ref{smallshift}) for $^2{\rm H}$ and $^{3}{\rm He}$ and 
eq.~(\ref{li7}) for $^{7}{\rm Li}$) between the various elemental abundances and 
the parameters $\eta$, $\delta G_{\rm N,f}$ and $\delta G_{\rm N,d}$.
In order to check their validity, we compare them with 
the results of numerical calculations. 
As a first step, we consider the case of a
constant variation of $G_{\rm N}$ during the entire
period relevant for BBN
(i.e. $\delta G_{\rm N,f}=\delta G_{\rm N,d}\equiv\delta G_{\rm N}$). 
In this assumption, a linear fit to the numerical result 
gives for the $^{4}{\rm He}$ abundance: 
\begin{equation}\label{numerical}
\delta Y_{4} \simeq 0.35 ~\delta G_{\rm N} + 0.09 ~\log(\eta/\eta_{\rm CMB}) 
\end{equation}
with an accuracy at the level of $2\%$ or better 
in the range $\delta G_{\rm N}=0.75-1.25$ and 
$\eta=2~10^{-10}-10^{-9}$, in reasonable agreement 
with estimate~(\ref{Yp}).
For $^2{\rm H}$ and $^3{\rm He}$ , expanding around $\eta_{0}=\eta_{\rm CMB}$, one obtains:
\begin{eqnarray}\label{dnum}
\delta Y_{2} &\simeq&  
\gamma_{2}(\eta_{\rm CMB}) \left( \log (\eta/\eta_{\rm CMB}) - 
0.25 \,\delta G_{\rm N} \right) \\
\delta Y_{3} &\simeq&  
\gamma_{3}(\eta_{\rm CMB}) \left( \log (\eta/\eta_{\rm CMB}) 
- 0.24 \,\delta G_{\rm N} \right)\label{he3num}
\end{eqnarray}
in good agreement with predictions (\ref{smallshift}), 
with $\gamma_{2}(\eta_{\rm CMB})= -3.7$ and $\gamma_{3}(\eta_{\rm CMB})= -1.3$. 
In addition, the $^{7}{\rm Li}$ behavior can be described by:
\begin{equation}\label{li7num}
\delta Y_{7} \simeq 0.32~\delta G_{\rm N}+
 \gamma_{7}(\eta_{\rm CMB}) \left( \log (\eta/\eta_{\rm CMB}) 
-0.22 \,\delta G_{\rm N} \right)  
\end{equation}
as predicted by eq.~({\ref{li7}), with $\gamma_{7}(\eta_{\rm CMB})= 4.8$.\footnote{
We remark that, for $^2{\rm H}$, $^3{\rm He}$ and $^{7}{\rm Li}$, 
expanding around an arbitrary value $\eta_{0}$ in the range 
$\eta_{0}= 3\cdot10^{-10}-10^{-9}$ (and using the proper values $\gamma_{i}(\eta_{0})$)
one obtains the same results as those described by relations (\ref{dnum}), 
(\ref{he3num}) and (\ref{li7num}), with essentially the same numerical coefficients.}

\section{Response functions}

As underlined in the previous section, the production of each element {\it responds}
in its own way to a variation $\delta G_{\rm N}$ of the Newton constant.
For example, a change of $G_{\rm N}$ at the time of weak 
interaction freeze-out would have important consequences on the observed helium
abundance, giving instead negligible corrections to that of deuterium.
So far we have implicitly assumed that $G_{\rm N}(T)$ stays constant
(at a value not necessarily equal to the present one) during BBN,
or that it takes two different values at the two key epochs, marked as
$T_{\rm f}$ and $T_{\rm d}$. 
A more general analysis, that can account for a time dependence of
$G_{\rm N}(T)$ along the all BBN period, requires the introduction of 
suitable functions, which describe the response of each elemental abundance to
an arbitrary time-dependent modification of the early universe expansion rate.

We have determined 
numerically the {\it response functions}: $\varrho_{i}(\eta,T)$, which are defined by: 
\begin{equation}
\delta Y_{i}(\eta,\delta H(T)) = 
2 \int \varrho_{i}(\eta,T)\, \delta H(T) \,\,\frac{dT}{T}~,
\end{equation}
where $i=2,3,4$ and 7 and $\delta H(T)$ is the fractional variation (assumed to be small) 
of the expansion rate of the universe at the temperature $T$ with respect to its standard value.
We remark that, in the assumption that $G_{\rm N}(T)$ is slowly varying, one has
that $\delta G_{\rm N}= 2 \, \delta H(T)$ from eq.(\ref{H}). The above 
equation can then be simply rewritten as
\begin{equation}
\delta Y_{i}(\eta,\delta G_{\rm N}(T)) = 
\int \varrho_{i}(\eta,T)\, \delta G_{\rm N}(T) \,\,\frac{dT}{T}~,
\end{equation}
which shows that
$\varrho_{i}(\eta,T)$ is basically the functional derivative of 
$\ln Y_{i}(\eta,\delta G_{\rm N}(T))$ with respect to $\delta G_{\rm N}(T)$.}

The response functions allows us to identify unambiguously the 
key epochs for the production of the various elements and to emphasize that
different elements are sensitive to the value of the Newton constant at slightly
different times.
Our results, calculated for $\eta = \eta_{\rm CMB}$, are shown in Fig.~1. 
As expected, the functions
$\varrho_{i}(\eta,T)$ have two peaks corresponding to the weak interaction freeze-out
and to the epoch, just after the deuterium bottleneck, during which the various elements
are effectively synthesized. 
The width of the two peaks reflects the fact that the 
above processes are not instantaneous.
 One can essentially identify 
the range $\Delta T_{\rm f} = (0.2 - 2){\rm MeV}$ with the weak interaction 
freeze-out epoch and $\Delta T_{\rm d} = (0.02 - 0.2){\rm MeV}$ with the
various elements synthesis period.
The behavior of the functions $\varrho_{i}(\eta,T)$ also allows to give a more
quantitative meaning to the parameters $G_{\rm N,f}$ and $G_{\rm N,d}$, which have
to be intended, evidently, as the average values of the Newton constant during the 
periods $\Delta T_{\rm f}$ and $\Delta T_{\rm d}$, respectively.

\begin{figure}[t]
\par
\begin{center}
\includegraphics[width=6.7cm,angle=0]{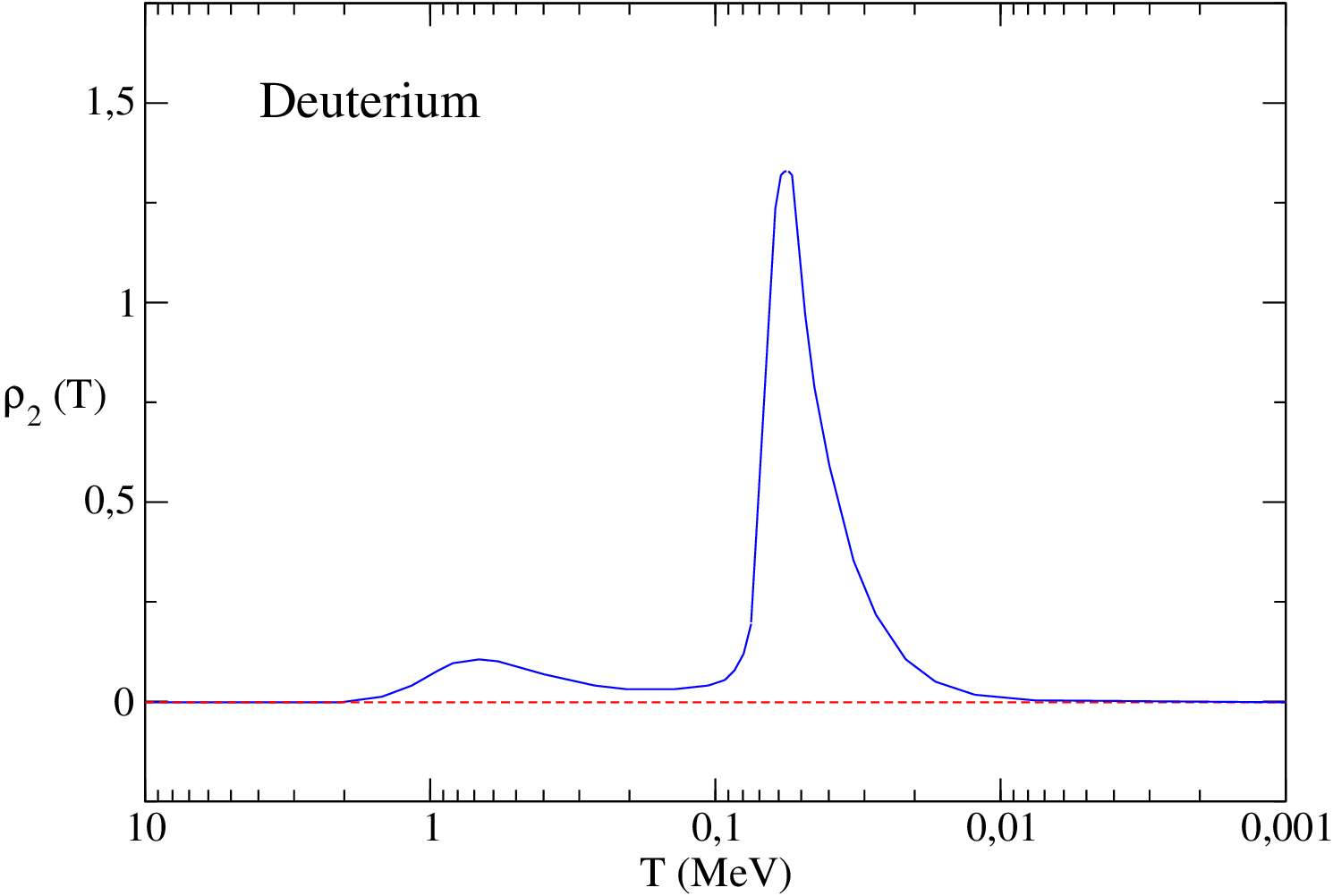}
\includegraphics[width=6.7cm,angle=0]{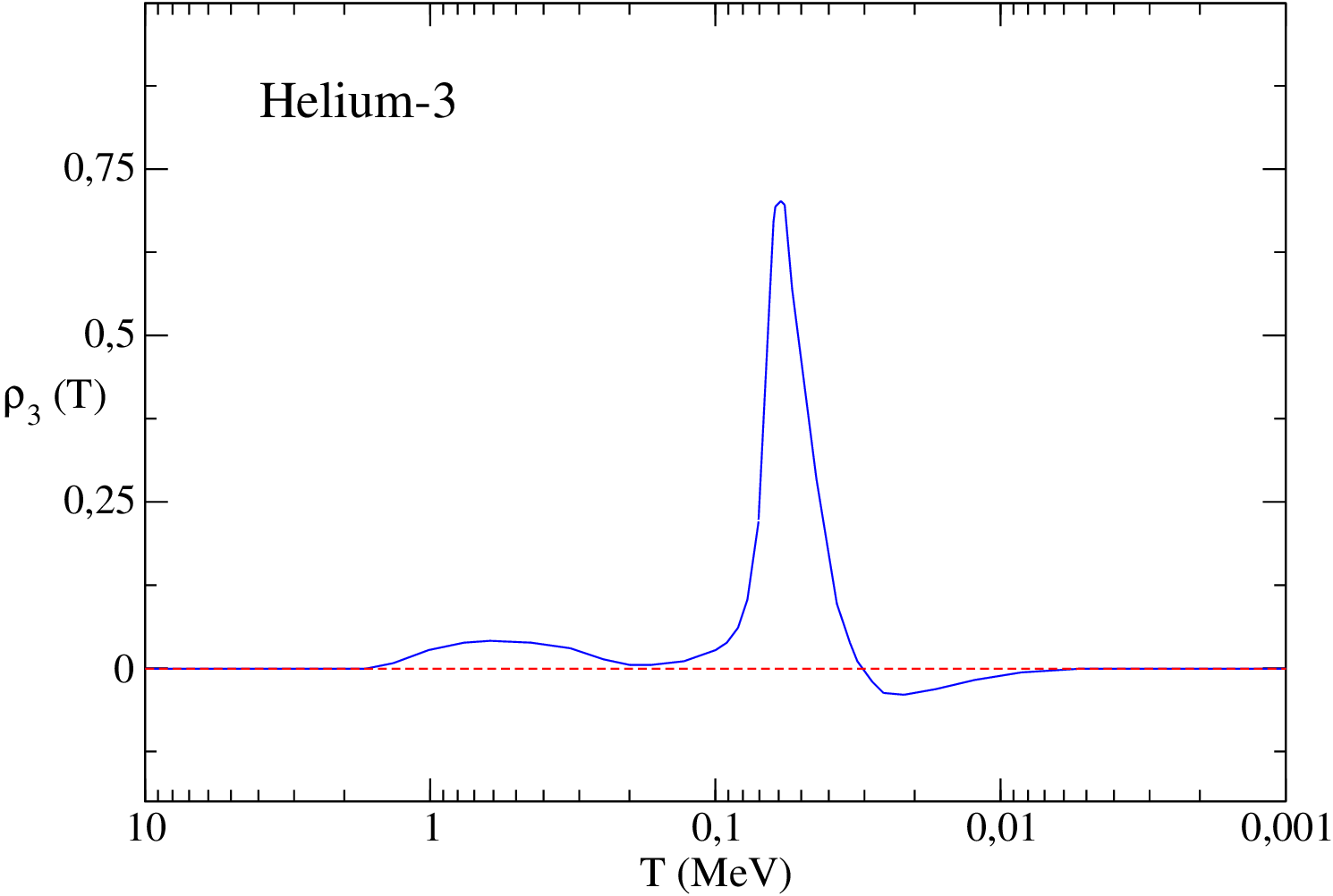}\\
\vspace{0.5cm}
\includegraphics[width=6.7cm,angle=0]{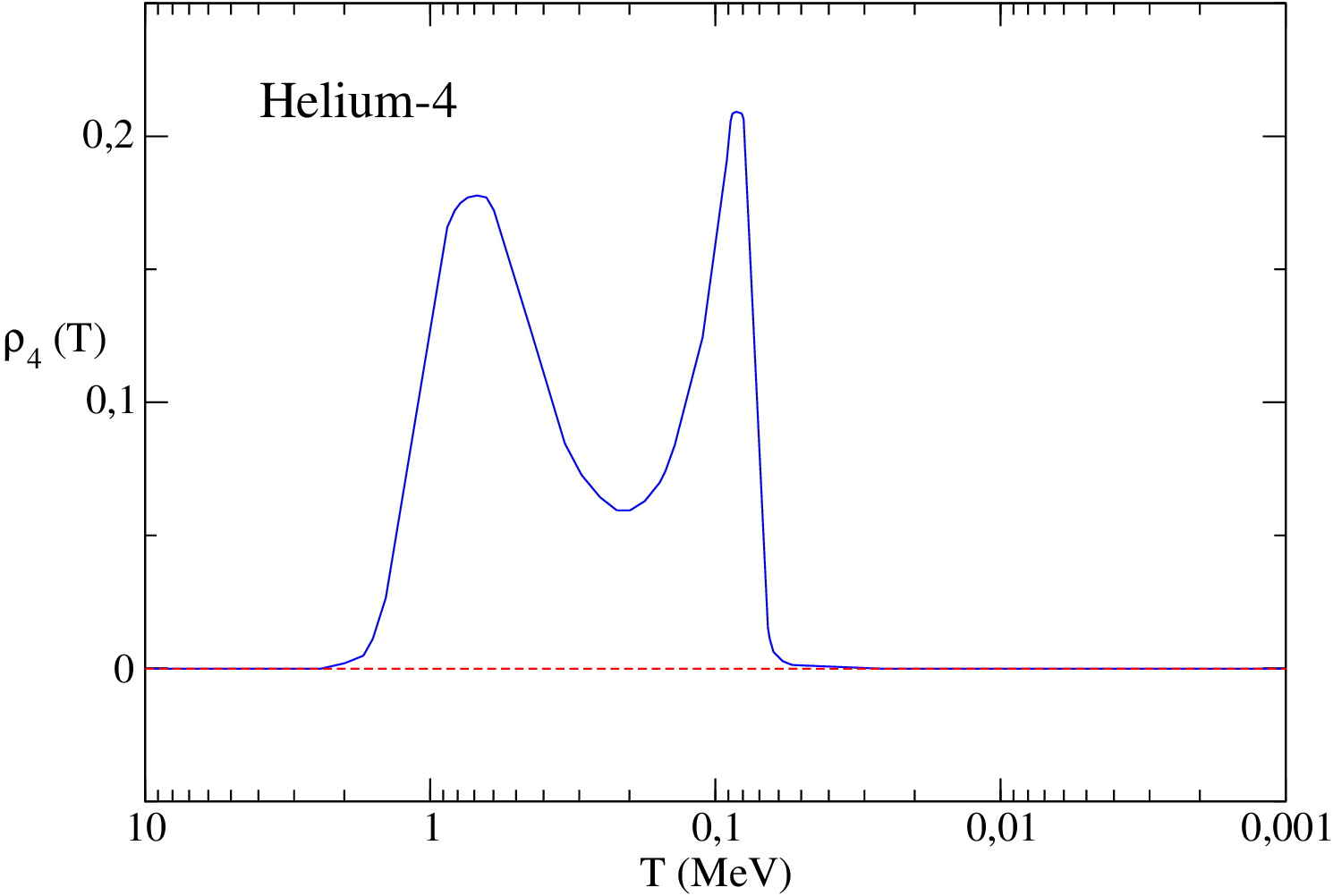}
\includegraphics[width=6.7cm,angle=0]{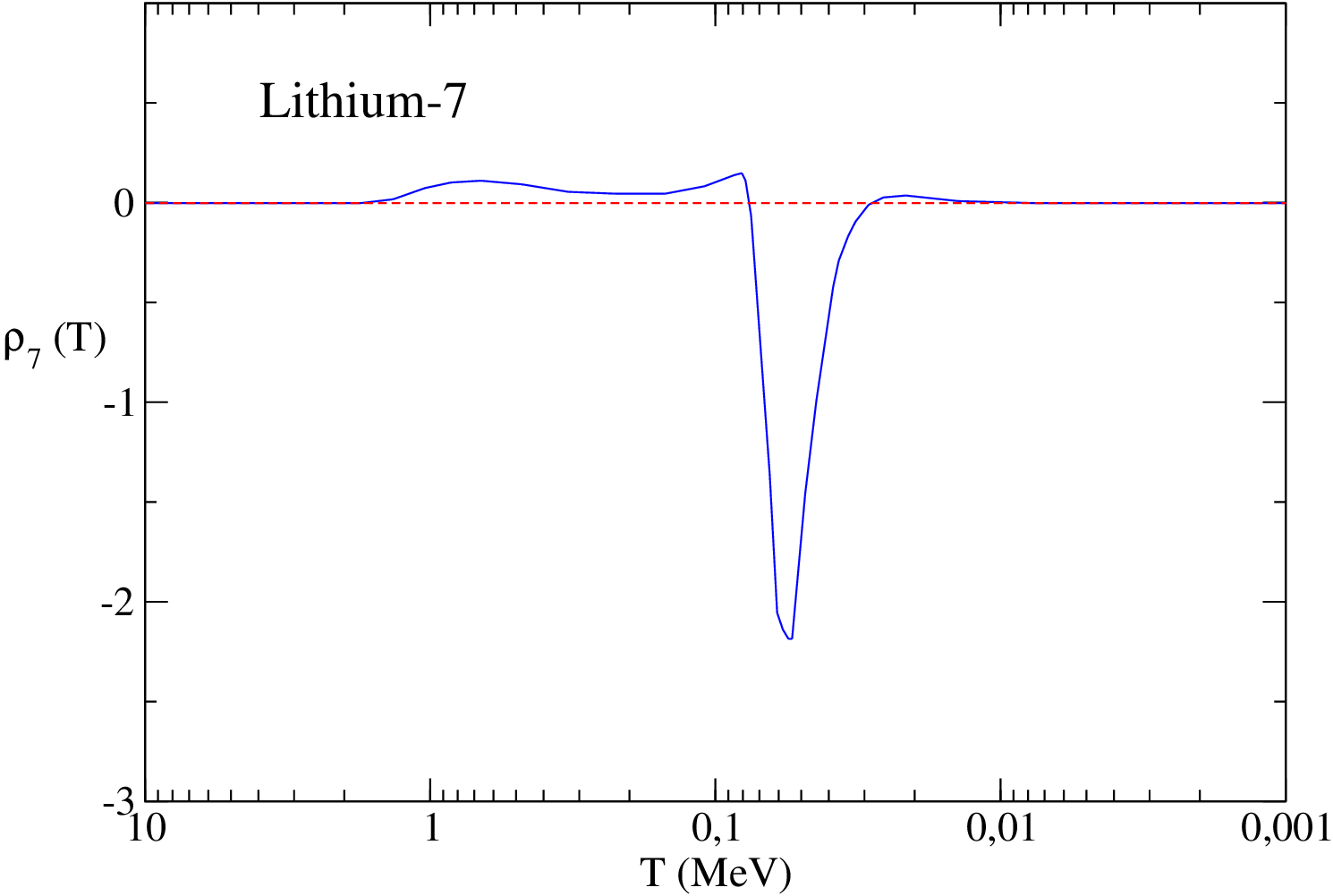}
\end{center}
\par
\vspace{-5mm} \caption{{\protect\small The response functions $\varrho_{i}(\eta,T)$ 
as a function of the temperature $T$, for $\eta=\eta_{\rm CMB}$. 
The functions $\varrho_{i}(\eta,T)$ describe the effect 
of an arbitrary time-dependent modification of the early universe expansion rate on
the various elemental abundances (see text for details).}}
\label{fig:FIT}
\end{figure}


The total area under the curves in fig.~1 (integrated in $\ln T$) gives the
numerical coefficient $\delta Y_{\rm i}/\delta G_{\rm N}$, which 
are obtained in eqs.(\ref{numerical}-\ref{li7num}) in the assumption of constant $G_{\rm N}$ variations. 
It is interesting, however, to consider separately the early time and the 
late time behavior of the functions $\varrho_{i}(\eta,T)$ in order
to have a feeling of the relative importance of the different epochs in
the various elements  production. In this spirit, we have calculated the 
numerical values:
\begin{eqnarray}
\alpha_{i} &=& \int_{0.2 {\rm MeV}}^{2 {\rm MeV}} \varrho_{i}(\eta,T) \,\,\frac{dT}{T} \\
\beta_{i} &=& \int_{0.02 {\rm MeV}}^{0.2 {\rm MeV}} \varrho_{i}(\eta,T)\,\,\frac{dT}{T}
\end{eqnarray}
for $i=2,3,4$ and 7, which have to be compared with the coefficients 
$\delta Y_{i}/\delta G_{\rm N,f} $ and $ \delta Y_{i}/\delta G_{\rm N,d}$
estimated in eqs.(\ref{Yp},\ref{smallshift},\ref{li7}). 
For $\eta= \eta_{\rm CMB}$, one obtains 
$\alpha_2 = 0.12 $ and $\beta_2 = 0.80$ for deuterium,
$\alpha_3 = 0.04$ and $\beta_3=0.29$ for helium-3,
$\alpha_4 = 0.22$ and $\beta_4=0.12$ for helium-4 and
$\alpha_7 = 0.14$ and $\beta_7=-0.83$ for lithium-7,
in good agreement with our predicted values.

One sees that $^{2}{\rm H}$, $^{3}{\rm He}$ and $^{7}{\rm Li}$ 
abundances are essentially determined by
the expansion rate of the universe during and after the d-bottleneck.
As a consequence, the bounds obtained from these elements
are basically bounds on $\delta G_{\rm N,d}$.
Helium-4, instead, is mainly sensitive to the value of the Newton constant during
the weak interaction freeze-out. However, the ``response'' function $\varrho_{4}(T,\eta)$ 
is rather broad, showing that $^4{\rm He}$ is sensitive to a rather long 
period, $0.05 \le T \le 2 ~{\rm MeV}$, of the early universe evolution.
In terms, of $G_{\rm N,f}$ and $G_{\rm N,d}$, the bounds obtained
from $^{4}{\rm He}$ observational data can be considered limits on the
combination $0.65 \delta G_{\rm N,f} + 0.35 \delta G_{\rm N,d}$.

Finally, we remark that the response functions $\varrho_{i}(\eta,T)$ have a 
quite general meaning and may be easily applied to discuss any non-standard 
schemes (e.g. new light - stable or decaying - particles, non vanishing muon or tau neutrino chemical potentials, etc.) whose main effect is to modify the early universe expansion rate.

\section{The BBN bound on $\delta G_{\rm N}$}

By comparing theoretical 
predictions with observational data for light element primordial abundances
one is able, in principle, to obtain bounds for $\delta G_{\rm N,f}$ and 
$\delta G_{\rm N,d}$.
However, comparison of theoretical results with observational data is not
straightforward because the data are subject to poorly known systematic errors and
evolutionary effects (see \cite{steigman} for a review).
The present situation can be summarized 
as it follows:\\
{\it i)} 
Recent determinations of deuterium in quasar absorption line systems (QAS) 
report values of ${\rm D/H}$ 
in the range ${\rm D/H} \sim 2-4 \times 10^{-5}$.  
However, the dispersion among the  
different determinations is not consistent with 
errors in the single measurements.  
We will use, in the following, the value  
${\rm D/H}=2.78^{+0.44}_{-0.38}\times 10^{-5}$  
given in~\cite{d7}, which is the weighted mean of
most recent deuterium determinations (see~\cite{d7} for 
detailed discussion and references).\\
{\it ii)} Independent determinations of $^{4}{\rm He}$ primordial abundance 
have statistical errors at the level of $1-2\%$ but  
differ among each others by about $\sim 5\%$. In particular, by 
using independent data sets, Olive et al. \cite{olive-he1,olive-he2}
have obtained $Y_{4} = 0.234 \pm 0.003$,  
while Izotov et al. \cite{izotov-he1,izotov-he2}
have found $Y_{4} = 0.244 \pm 0.002$. 
 We will use the ``average'' value $Y_{4}=0.238$, quoted in \cite{olive-he3}, 
with the error estimate $\Delta Y_{4} =0.005$, which is  
obtained from the dispersion of the various $Y_{4}$ determinations. \\
{\it iii)} The $^{7}{\rm Li}$ and $^{3}{\rm He}$ primordial abundances are not 
known, at present, with a level of uncertainty comparable to the other elements.
The reported values for the primordial abundances (see \cite{litio7} for 
$^{7}{\rm Li}$ and \cite{bania} for $^{3}{\rm He}$) are, evidently, important as 
a confirmation of the BBN paradigm, but are presently not very effective 
in constraining possible non-standard 
BBN scenarios. For this reason we will not include these elements in our analysis.

In Fig.~2  we discuss the bounds on $\delta G_{\rm N}$ and $\eta$ that can be obtained 
by comparing theoretical predictions with observational data for primordial $^{2}{\rm H}$ 
and $^4{\rm He}$. 
Theoretical calculations are made in the assumption of a constant variation, 
$\delta G_{\rm N}$, of the Newton constant in the period relevant for BBN.
This simple assumption is motivated by the fact that the present observational
situation does not allow to determine the evolution of 
$G_{\rm N}(t)$ {\it during} BBN.
However, when considering a theoretical framework 
in which a specific time-dependence for $G_{\rm N}(t)$ is predicted, 
one has to keep in mind that $^{2}{\rm H}$ and $^{4}{\rm He}$ respond 
differently to a non-constant modification of the early universe expansion rate.

The results shown in Fig.~2 are obtained by defining a $\chi^2(\eta,\delta G_{\rm N})$
as prescribed in \cite{noi} which takes into account both observational 
and theoretical errors in the various elemental abundances. The best fit points
in the plane $(\log \eta, \delta G_{\rm N})$ are obtained by minimizing the
$\chi^2$. 
The three confidence level (C.L.) curves (solid, dashed, and dotted)
are defined by $\chi^{2}-\chi^{2}_{\rm min}=2.3,\, 6.2,\, 11.8$, corresponding to 
68.3\%, 95.4\% and 99.7\% C.L. for two degrees of freedom ($\eta$ and $\delta G_{\rm N}$),
i.e. to the probability intervals designated as 1, 2, and 3 standard deviation limits.
The upper panels are obtained by considering BBN alone, while the lower panels 
show the bounds which can be obtained by combining BBN data
with the measurement of the baryon to photon ratio from CMB and LSS. 
This is done by adding the contribution:
\begin{equation}
\chi^2_{\rm CMB}(\eta) = \frac{(\eta - \eta_{\rm CMB})^2}{\sigma_{\rm CMB}^2}
\end{equation}
to the BBN chi-square, where $\eta_{\rm CMB}=6.14\cdot 10^{-10}$ is the baryon-to-photo ratio 
determined by CMB and LSS data and $\sigma_{\rm CMB}=0.25\cdot 10^{-10}$ is the error in this 
determination \cite{spergel}. 
In this case, the C.L. curves are determined by the condition 
$\chi^{2}-\chi^{2}_{\rm min}=1,\, 4,\, 9 $, corresponding to 
1, 2, and 3 standard deviations for one degree of freedom ($\delta G_{\rm N}$),
 since $\eta$ is considered as a measured quantity
 \footnote{Technically speaking, one should define $\chi^{2}_{\rm tot}(\delta G_{\rm N})
=\min_{\eta} [\chi^{2}(\eta,\delta G_{\rm N})+\chi^{2}_{\rm CMB}(\eta)]$ 
which depends only on the parameter $\delta G_{\rm N}$. For graphical reasons and to 
facilitate the comparison with the bounds obtained only from BBN, we presented 
the results in the two dimensional plane $(\eta,\delta G_{\rm N})$.}. 
 
 The results displayed in fig.~2 allow to obtain the following conclusions: \\
{\it i)} The $^{2}{\rm H}$ and $^4{\rm He}$ data (panels $\it a$ and $\it b$)
select bands in the plane $(\log(\eta),\delta G_{\rm N})$, which can be easily
interpreted in terms of the analitycal (eqs.(\ref{Yp}) and (\ref{smallshift})) 
and numerical (eqs.(\ref{numerical}) and (\ref{dnum})) relations discussed in the previous section.
In order to have a bound on $\delta G_{\rm N}$ one has to combine 
the $^{2}{\rm H}$ and $^{4}{\rm He}$ observational results (panel $\it c$)
(in the assumption that $\delta G_{\rm N}(t)$ stays nearly constant 
during BBN) and/or to consider the independent information 
on $\eta$ given by CMB+LSS (lower panels).\\
{\it ii)} If we combine $^{2}{\rm H}$ and CMB+LSS data (panel $\it d$), 
we obtain $\delta G_{\rm N} = 0.09 ^{+0.22} _{-0.19}$, in agreement with \cite{krauss}.
The quoted bound is consistent with the standard 
assumption  that $G_{\rm N}$ has not varied during the evolution of the universe
and is, at present stage, the most robust piece of information
on the value of the gravitational constant in the early universe.
We recall that this bound essentially applies to the value of the Newton constant 
during and after the d-bottleneck (i.e. when $0.02 {\rm MeV}\le T \le 0.2 {\rm MeV})$
or, equivalently (in the previous sections notations), 
to the parameter $\delta G_{\rm N,d}$.\\
{\it iii)} If we combine the $^{4}{\rm He}$ and CMB+LSS data (panel $\it e$), 
we obtain $\delta G_{\rm N} = - 0.11 \pm 0.05$, which shows that 
$^{4}{\rm He}$ observational data favor a reduction of $G_{\rm N}$ in the early 
universe, even if errors are large enough to allow for 
the standard value $\delta G_{\rm N}=0$. 
This result is emphasized (reduced) if we consider the ``low'' 
(``high'') helium value $Y_{4}=0.234\pm0.003$ given in \cite {olive-he1,olive-he2}
($Y_{4}=0.244\pm 0.002$ given in \cite{izotov-he1,izotov-he2}),
which results in $\delta G_{\rm N}= -0.15 \pm 0.03$ 
($\delta G_{\rm N}= - 0.05 \pm 0.02$) and is also obtained in 
panels {\it c)} and {\it f)} where $^{2}{\rm H}$ observational data
are also included. If confirmed (and strengthened) by future data, 
this could be an important indication of non-standard
effects in BBN. We remark, however, that the present situation is quite delicate. 
The uncertainty in the quoted bound, $\delta G_{\rm N}=-0.11\pm0.05$, is completely 
dominated by (not well known) systematic errors in $^{4}{\rm He}$ measurements. 
It is thus advisable to wait for a better comprehension of these errors, 
before a final result can be obtained.\\
{\it iv)} In panel {\it f} we show the bound, $\delta G_{\rm N}=-0.09\pm0.05$, 
that is obtained by considering $^{2}{\rm H}$, $^{4}{\rm He}$ and CMB+LSS data. 
We see that the fit is dominated by helium-4 and CMB+LSS observational data, 
and that the information provided by deuterium only marginally reduces the
error bar with respect to the previous case.
We remark that the above result is obtained in the assumption that $\delta G_{\rm N}$ 
is nearly constant during BBN. 
In principle, one could fit the data by considering $\delta G_{\rm N,f}$
and $\delta G_{\rm N,d}$ as independent parameters. In a future perspective, 
this could be interesting as a possible test for variations of $G_{\rm N}$ 
{\it during} BBN (or equivalent schemes). 
In the present experimental situation this appears too ambitious. 
The value of $\chi^{2}_{\rm min}=1.0$ for 
the best-fit point indicate, in fact, that 
the quality of the fit is good and that there is no real evidence, 
at present, in favor of theoretical schemes which predict 
non-constant modification  of the early universe expansion rate.

\begin{figure}[t]
\par
\begin{center}
\includegraphics[width=4.9cm,angle=0]{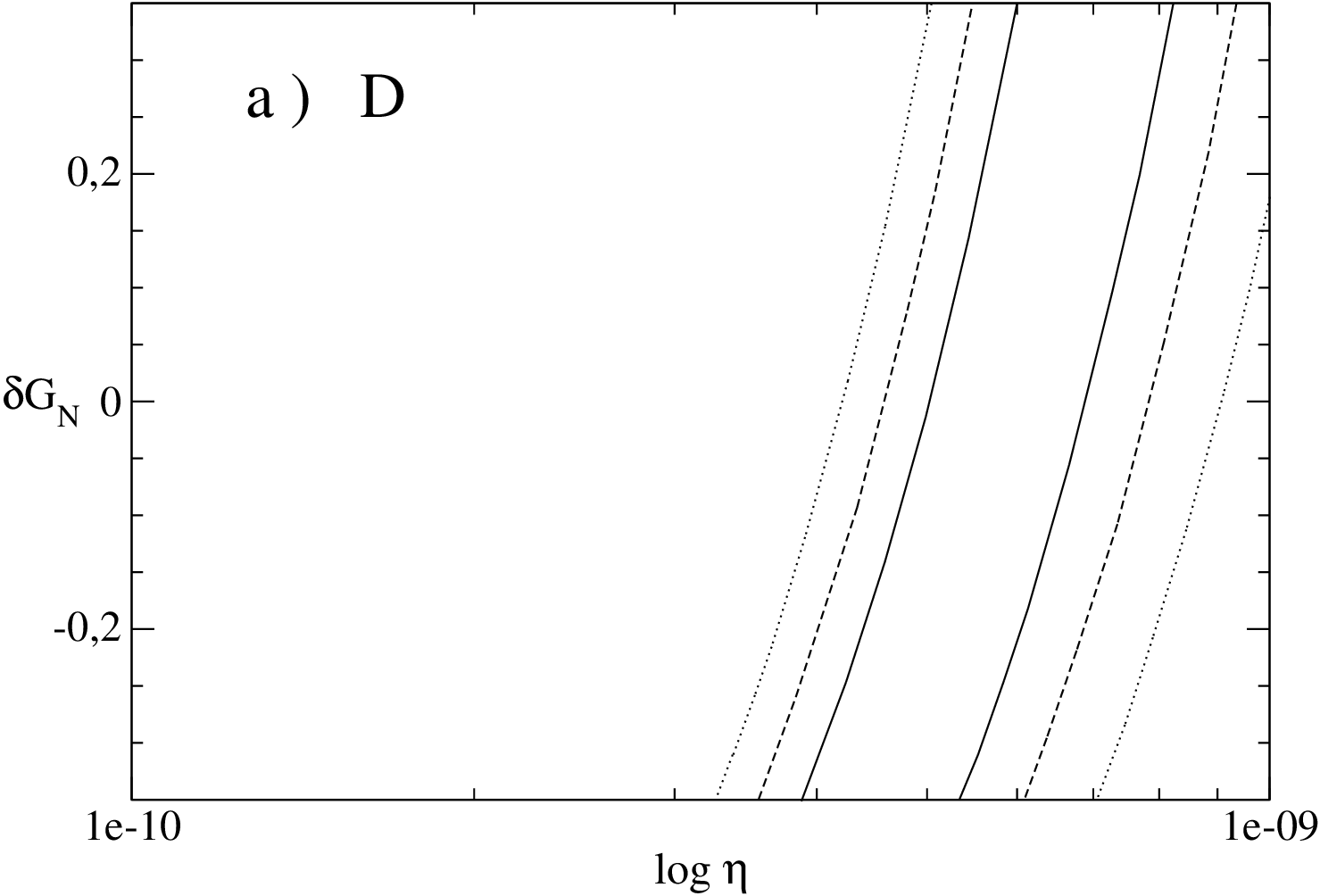}
\includegraphics[width=4.9cm,angle=0]{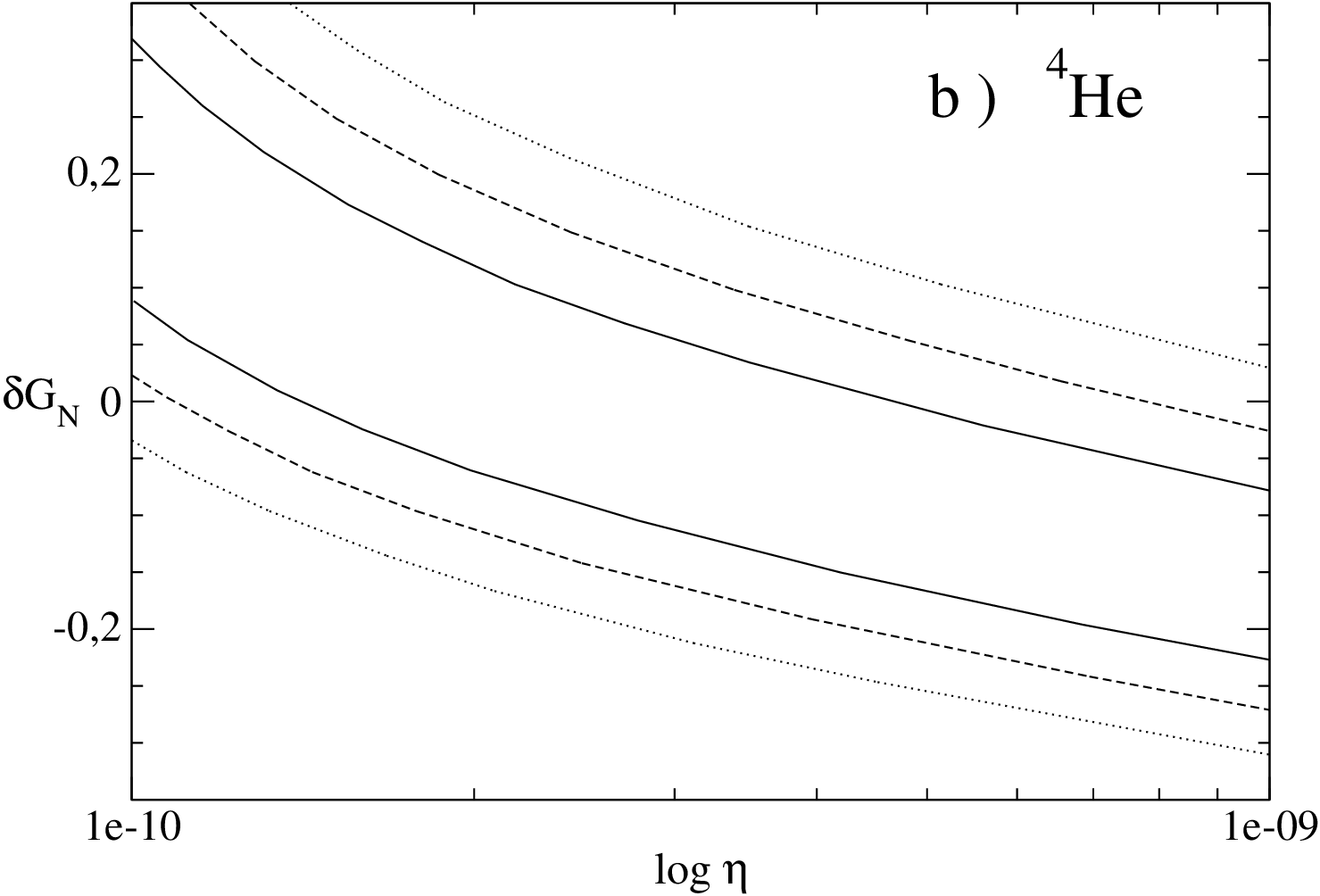}
\includegraphics[width=4.9cm,angle=0]{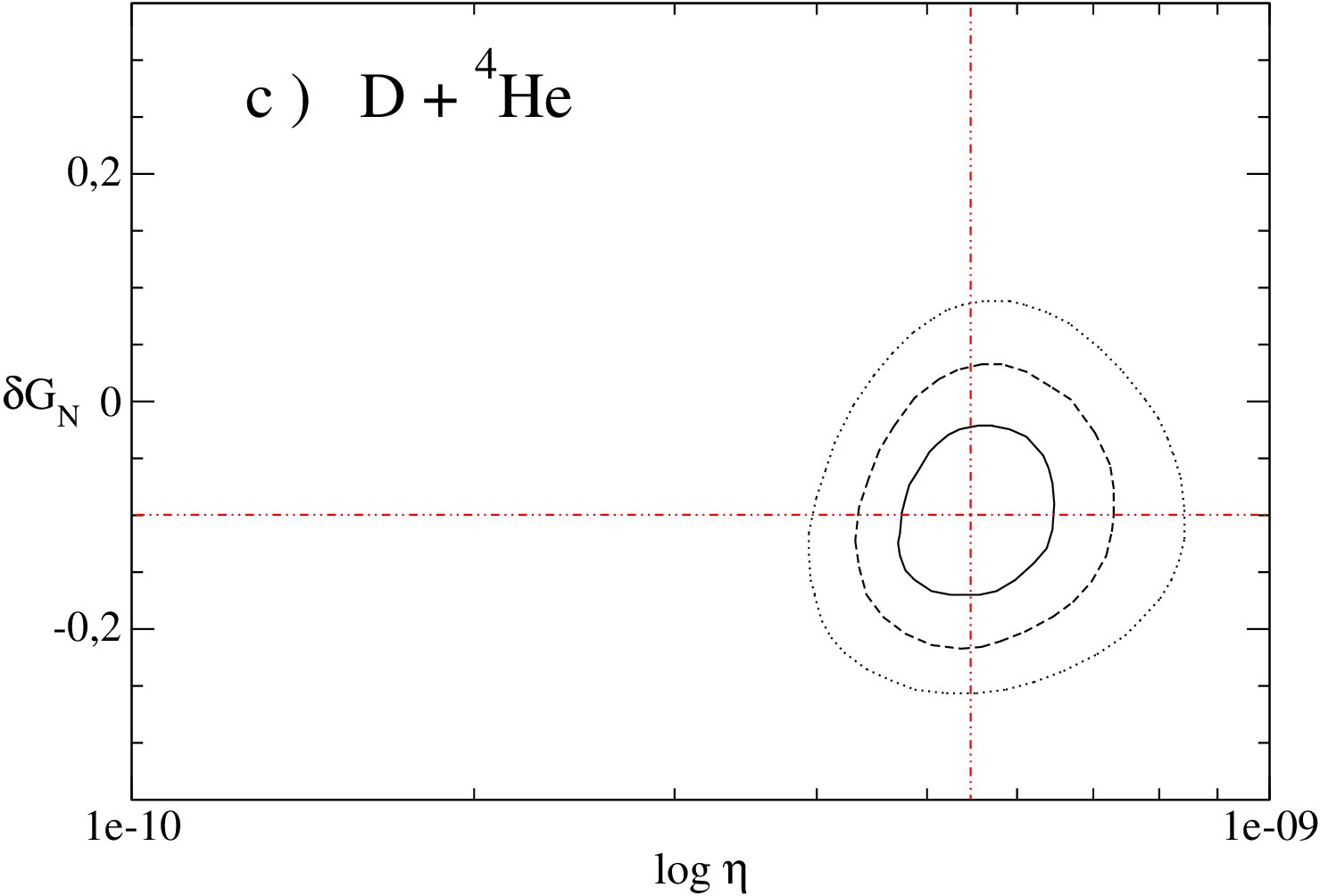}\\
\vspace{0.5cm}
\includegraphics[width=4.9cm,angle=0]{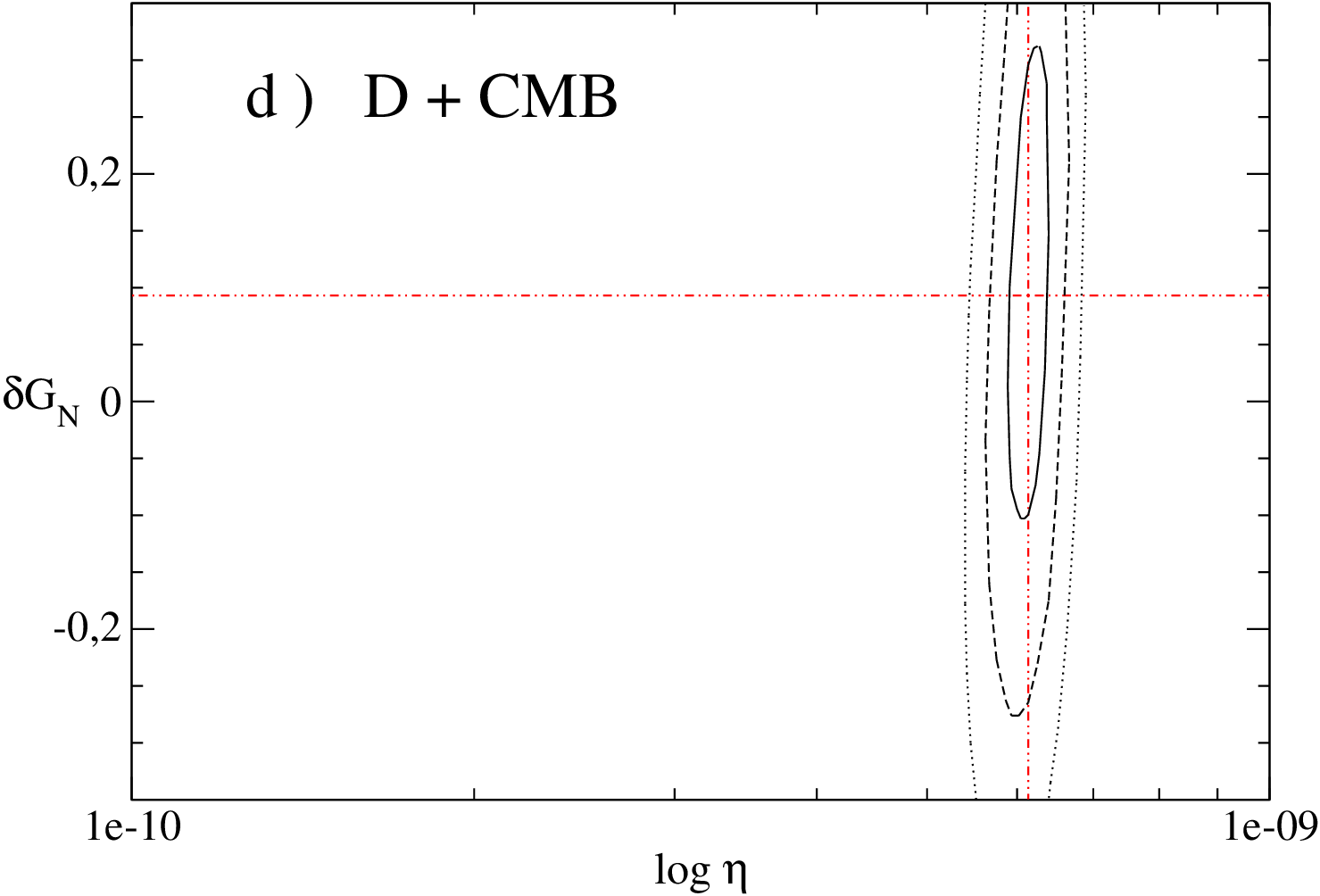}
\includegraphics[width=4.9cm,angle=0]{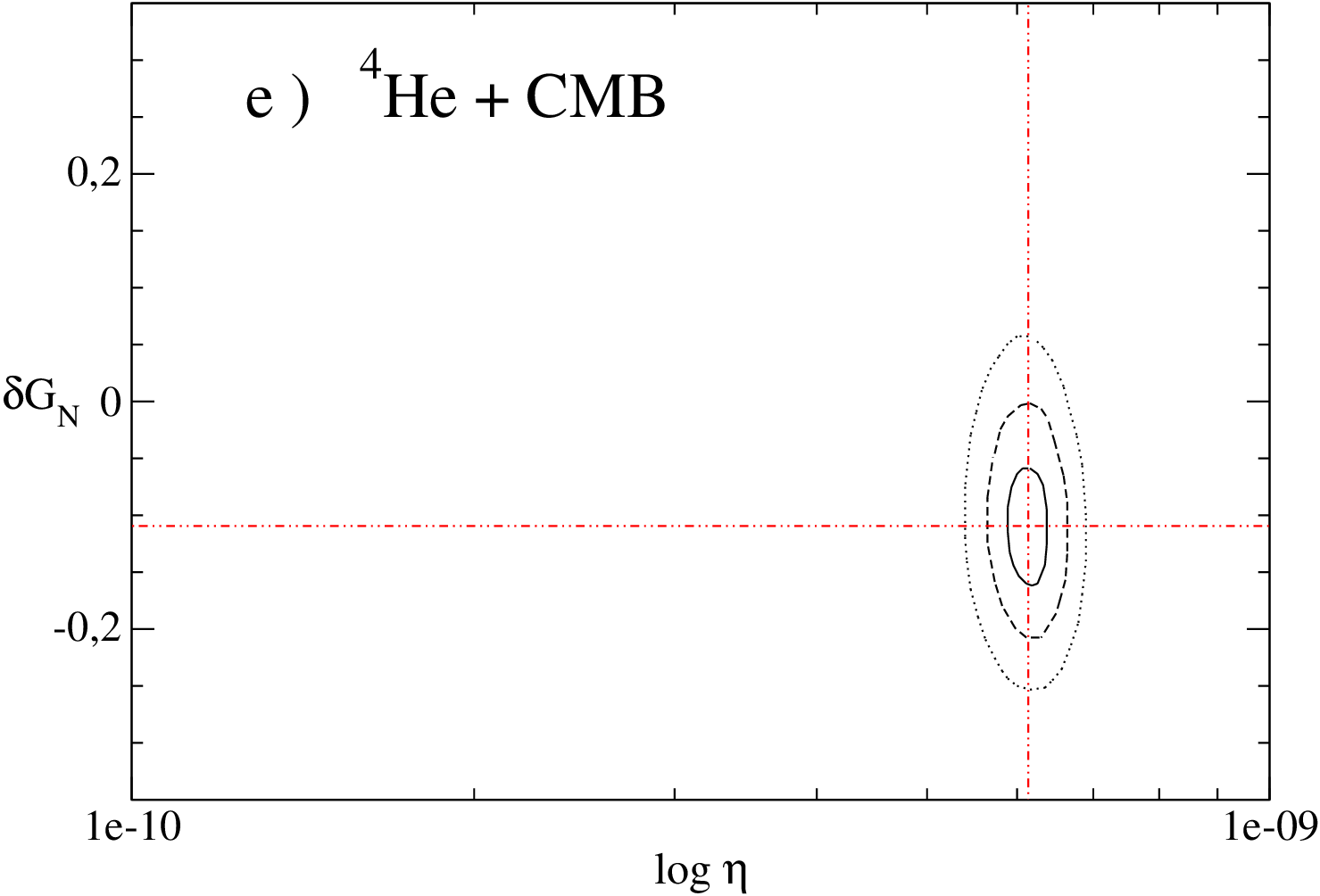}
\includegraphics[width=4.9cm,angle=0]{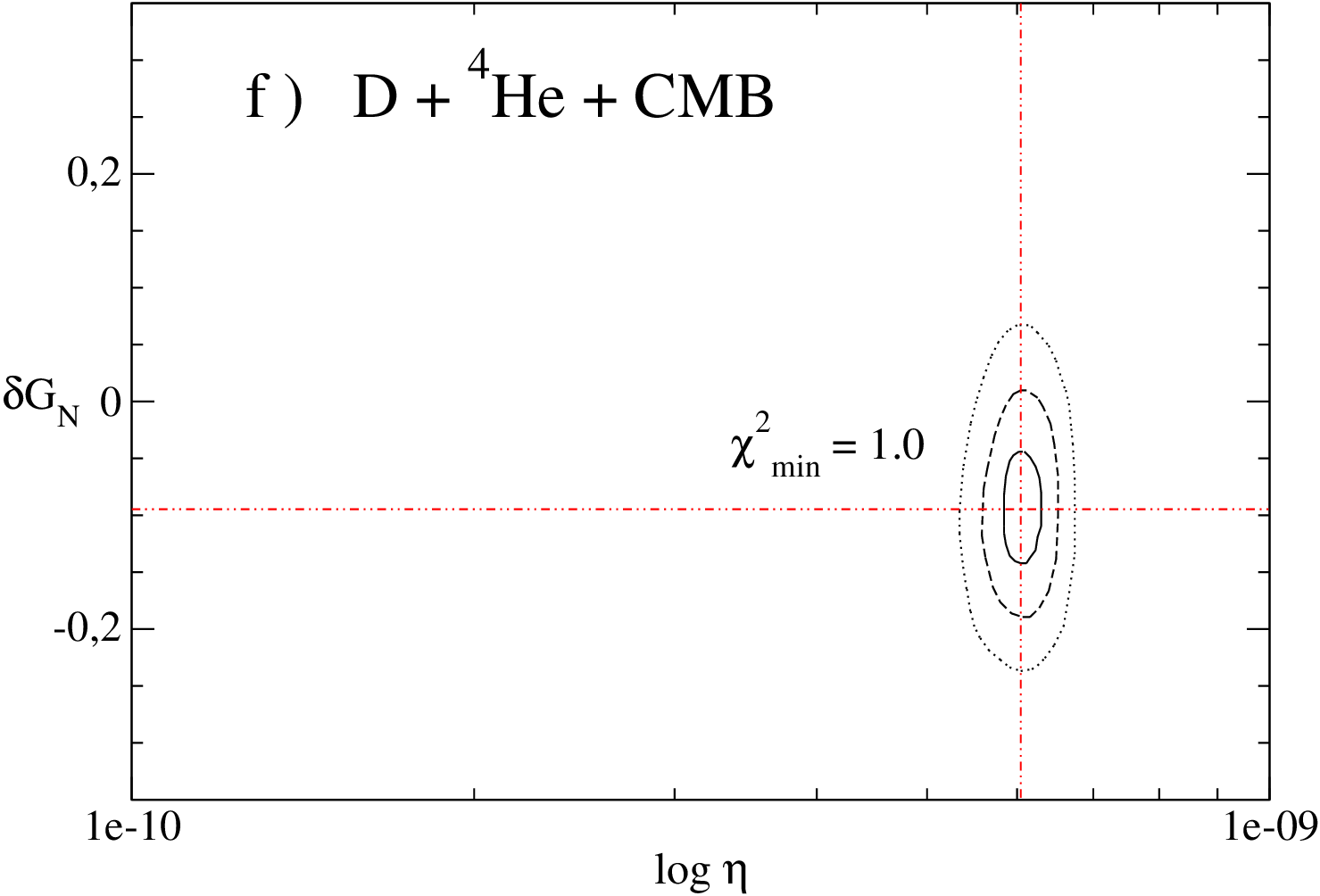}
\end{center}
\par
\vspace{-5mm} \caption{{\protect\small 
The bounds on $\delta G_{\rm N}$
which can be obtained from $^2{\rm H}$ and $^4{\rm He}$ observational data.
The upper panels are obtained considering BBN alone. The lower panels 
are obtained by combining BBN data with the measurement of the baryon to photon 
ratio from CMB and LSS (see text for details).}}
\label{fig:BOUNDS }
\end{figure}


\section{Summary and conclusions}
We summarize the main points of this letter and provide some perspective:\\
{\it i)} We have discussed the role of $G_{\rm N}$ in BBN and we have 
derived analytically the dependence of light element 
primordial abundances on the values ($G_{\rm N,f}$ and $G_{\rm N,d}$) 
of the Newton constant at the key epochs:
the weak interaction freeze-out epoch and the epoch, just after the deuterium
bottleneck, during which the elements are effectively synthesized.\\
{\it ii)} 
The production of each element {\it responds}
in its own way to a variation $\delta G_{\rm N}$ of the Newton constant.
A general study, that can account for a time dependence of $G_{\rm N}(T)$
during the BBN period, requires the introduction of suitable functions which
describe the response of each element to an arbitrary time-dependent modification
of the early universe expansion rate. We have numerically
calculated these response functions, 
obtaining a good agreement with the analytical estimates.\\
{\it iii)}
We have emphasized
that different elements are sensitive to the value of the Newton constant at
slightly different times. The $^{2}{\rm H}$, $^{3}{\rm He}$ and $^{7}{\rm Li}$ 
abundances are essentially determined by
the expansion rate of the universe close to the d-bottleneck.
Helium-4 is, instead, mainly sensitive to the value of the Newton constant during
the weak interaction freeze-out. In a future perspective, this could be interesting as a possible test for 
variations of $G_{\rm N}$ during BBN (or equivalent schemes).\\
{\it iii)} 
We have discussed the observational bounds on the possible 
variations of the gravitational constant in the early universe. 
Our best limit, $\delta G_{\rm N}=0.09^{+0.22}_{-0.19}$ , is obtained by combining $^{2}$H observational results with the measurements of the baryon to photon ratio obtained from
CMB and LSS data. This limit refers to the value of $G_{\rm N}$ when the temperature
of the universe is $0.02\le T \le 0.2$ MeV (i.e. during and immediately after the d-bottleneck epoch) and is consistent with the standard assumption that $G_{\rm N}$ has not varied during the evolution of the universe.\\
{\it iv)}
To conclude, we remark that the results obtained for $\delta G_{\rm N}$ may be easily
applied to a more general context in which other constants are allowed to vary and/or
new light particles are included. The limits on $\delta G_{\rm N,f}$ and $\delta G_{\rm N,d}$
are, indeed, essentially obtained by comparing the expansion rate of the Universe with 
the weak reaction rate (in the case of $\delta G_{\rm N,f}$) 
or with the light element nuclear reaction rates (in the case of $\delta G_{\rm N,d}$). 
One immediately understands, then, that the bound on $G_{\rm N,f}$ is basically a bound on 
$ g_{*}(T_{\rm f})\,G_{\rm N,f}/G_{\rm F}^{4}$, where $G_{\rm F}$ 
is the Fermi constant and $g_{*}(T_{\rm f})$ is the number 
of relativistic degrees of freedom at the weak interaction freeze-out, 
while the bound on $G_{\rm N,d}$ is essentially a bound on
$g_{*}(T_{\rm d}) \, G_{\rm N,d}$, where $g_{*}(T_{\rm d})$ is the number 
of relativistic degrees of freedom at the d-bottleneck.

\section*{Acknowledgments}\

We thank G. Fiorentini for useful discussions and for earlier collaboration 
on the subject of this paper. We are also grateful to A. D. Dolgov for useful
suggestions and comments.

\end{document}